\documentclass[twocolumn,showpacs,preprintnumbers,amsmath,amssymb,pra]{revtex4-1}

\usepackage{graphicx}
\usepackage{dcolumn}
\usepackage{bm}
\usepackage{color}

\begin{document}

\title{Bifurcation-based adiabatic quantum computation with a nonlinear oscillator network:
Toward quantum soft computing}

\author{Hayato Goto}
\affiliation{Frontier Research Laboratory, 
Corporate Research \& Development Center, 
Toshiba Corporation, 
1, Komukai Toshiba-cho, Saiwai-ku, Kawasaki-shi, 212-8582, Japan }

\begin{abstract}

The dynamics of nonlinear systems qualitatively change depending on their parameters, 
which is called bifurcation. 
A quantum-mechanical nonlinear oscillator can yield 
a quantum superposition of two oscillation states, 
known as a Schr\"odinger cat state, via quantum adiabatic evolution through its bifurcation point. 
Here we propose a quantum computer comprising such quantum nonlinear oscillators, 
instead of quantum bits, to solve hard combinatorial optimization problems. 
The nonlinear oscillator network finds optimal solutions via quantum adiabatic evolution, 
where nonlinear terms are increased slowly, 
in contrast to conventional adiabatic quantum computation or quantum annealing,
where quantum fluctuation terms are decreased slowly. 
As a result of numerical simulations, 
it is concluded that quantum superposition and quantum fluctuation work effectively 
to find optimal solutions. 
It is also notable that the present computer is analogous to neural computers, 
which are also networks of nonlinear components. 
Thus, the present scheme will open new possibilities for quantum computation, 
nonlinear science, and artificial intelligence.

\end{abstract}


\maketitle

Nonlinearity is the origin of various interesting phenomena, 
such as chaos, fractal, and bifurcation \cite{Strogatz}. 
A bifurcation is a parameter-dependent qualitative change in nonlinear dynamics, 
such as a transition from a single stable state to two stable ones (bistability). 
As a result of recent advances in nanotechnology, 
artificial nonlinear oscillators may possess both large nonlinearity and low loss simultaneously 
and consequently enter the quantum regime \cite{Dykman}. 
A remarkable example is the generation of a quantum superposition of two oscillation states, 
known as a Schr\"odinger cat state, 
with superconducting microwave resonators coupled to a superconducting artificial atom 
\cite{Leghtas2015a}, 
where nonlinear dissipation exceeds linear one. 
Although a scheme for quantum computation using such cat states as quantum bits (qubits) 
has been proposed
\cite{Mirrahimi2014a}, 
continuous degrees of freedom of quantum nonlinear oscillators have not been fully harnessed.

Here we first show that a quantum-mechanical nondissipative oscillator 
with desirable nonlinearity can yield a cat state via quantum adiabatic evolution 
through its bifurcation point. 
Next, we propose a quantum computer comprising such quantum nonlinear oscillators, 
which exploits a superposition of an exponentially large number of states 
of the nonlinear oscillator network to solve combinatorial optimization problems. 
The nonlinear oscillator network finds optimal solutions via quantum adiabatic evolution, 
as conventional adiabatic quantum computation \cite{Farhi2000a,Farhi2001a} 
or quantum annealing \cite{Kadowaki1998a,Santoro2002a,Das2008a} does. 
However, these mechanisms are different: 
whereas in quantum annealing quantum fluctuation terms are decreased slowly, 
in the present computation nonlinear terms are increased slowly. 
To distinguish them, we refer to the present approach 
as bifurcation-based adiabatic quantum computation. 
Finally, we present numerical simulation results indicating 
that quantum superposition and quantum fluctuation work effectively 
to find optimal solutions.

We start with a single quantum nonlinear oscillator. 
The oscillator used here is a parametrically driven Kerr (or Duffing) nonlinear oscillator (KPO). 
Interestingly, this is similar to a swing, 
where a pendulum (approximate Duffing oscillator) is driven 
by modulating the eigenfrequency by changing the height of the center of mass 
(parametric driving). 
This is not only the simplest one for the present purpose
but also physically feasible.
Promising candidates for implementing this model
are superconducting microwave resonators with Josephson junctions
\cite{Leghtas2015a,Vlastakis2013a,Kirchmair2013a,Rehak2014a,Lin2014a} (Chap. 15 in \cite{Dykman}),
nanoelectromechanical systems \cite{Okamoto2013a,Faust2013a} (Chap. 10 in \cite{Dykman}), 
and carbon nanotubes \cite{Moser2014a} (Chaps. 12 and 13 in \cite{Dykman}).

In a frame rotating at half the pump frequency of the parametric drive and in the rotating-wave approximation, 
its Hamiltonian is given by
\begin{align}
H_1=\hbar \Delta a^{\dagger} a + \hbar \frac{K}{2} a^{\dagger 2} a^2
- \hbar \frac{p}{2} (a^{\dagger 2} + a^2),
\label{eq-Hamiltonian1}
\end{align}
where 
$a$ and $a^{\dagger}$ are the annihilation and creation operators
for quanta of the oscillator 
(the quanta are, e.g., photons for electromagnetic resonators or phonons for mechanical oscillators),
$\Delta$ is the detuning of the oscillator eigenfrequency from half the pump frequency, 
$K$ is the Kerr coefficient for the Kerr effect,
and $p$ is the pump amplitude for the parametric drive \cite{Dykman}.
Hereafter,
we assume that $K$ and $\Delta$ are positive constants and
$p$ is a nonnegative control parameter. 
When $K$ is negative, similar discussion is straightforward.

Before describing the quantum dynamics,
we consider a classical model for the KPO.
This is helpful for understanding the quantum dynamics, as found below.
The classical equations of motion are 
\begin{align}
\dot{x} &= y \left[ \Delta + p + K \left( x^2+y^2 \right) \right],
\label{eq-classical-KPO-x}
\\
\dot{y} &= x \left[ -\Delta + p - K \left( x^2+y^2 \right) \right],
\label{eq-classical-KPO-y}
\end{align}
where the dots denote differentiation with respect to time $t$.
These equations are derived 
by replacing $a$ with a complex number $x+iy$ in the Heisenberg equation of motion
for $a$ with $H_1$.
Here $x$ and $y$ are real variables 
corresponding to the Hermitian operators $(a+a^{\dagger})/2$ and 
$(a-a^{\dagger})/2i$, respectively, often called quadrature amplitudes.

\begin{figure}[htbp]
	\includegraphics[width=8cm]{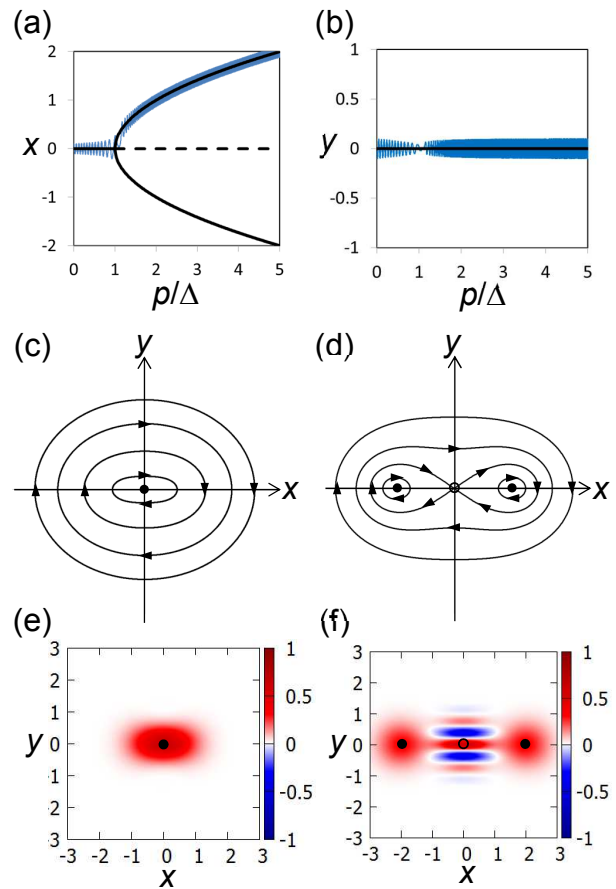}
	\caption{Dynamics of a single KPO. Here $\Delta$ is set to $K$.
	(a) and (b) show the bifurcation diagram (bold lines) 
	and the simulation result (thin lines) for the classical model, 
	where the solid and broken bold lines correspond to 
	the stable and unstable fixed points, respectively.
	(c) and (d) show the phase portraits for the classical model, where
	$p=0.9\Delta$ in (c) and $p=3\Delta$ in (d).
	(e) and (f) show the Wigner functions obtained by 
	numerically solving the Schr\"odinger equation with $H_1$ 
	in Eq. (\ref{eq-Hamiltonian1}), where
	$p=0.9\Delta$ in (e) and $p=5\Delta$ in (f).
	In (c)--(f), the filled and open circles represent 
	the classical stable and unstable fixed points, respectively.}
	\label{fig-single-oscillator}
\end{figure}

To grasp the dynamics of such a nonlinear system,
it is useful to investigate the fixed points,
which are defined by $\dot{x}=\dot{y}=0$ \cite{Strogatz}.
When $p \le \Delta$,
the origin is a single fixed point, which is stable.
When $p > \Delta$,
the origin becomes an unstable fixed point
and two stable ones are created,
the positions of which are $(\pm \sqrt{(p-\Delta)/K}$, 0).
Thus, the bifurcation point is at $p=\Delta$.
The dependence of the fixed points on $p$ is depicted 
in Figs. \ref{fig-single-oscillator}(a) and \ref{fig-single-oscillator}(b) 
as the bold lines, where 
the solid and broken lines correspond to the stable and unstable fixed points, respectively.
(In Fig. \ref{fig-single-oscillator}, $\Delta$ is set to $K$.)
Such figures are called bifurcation diagrams \cite{Strogatz}.

The oscillating thin curves in Figs. \ref{fig-single-oscillator}(a) and \ref{fig-single-oscillator}(b)
are obtained by numerically solving 
Eqs. (\ref{eq-classical-KPO-x}) and (\ref{eq-classical-KPO-y}),
where $p(t)$ is increased linearly from $p(0)=0$ to $p(500/K)= 5K$
and the initial condition is set as $x(0)=0.1$ and $y(0)=0$.
This result suggests that when the initial state is near a stable fixed point 
and the pump amplitude varies slowly,
the trajectory follows one of the stable branches of the bifurcation.

This result can be understood as follows.
This system with constant $p$ 
is conservative with the following conserved quantity:
\begin{align}
E(x,y)=
\frac{\Delta}{2} \left( x^2+y^2 \right)
+\frac{K}{4} \left( x^2+y^2 \right)^2 
- \frac{p}{2} \left( x^2-y^2 \right).
\nonumber
\end{align}
Thus, the trajectories are given by contours of $E(x,y)$ \cite{Strogatz}.
Examples of the trajectories are shown in Figs. \ref{fig-single-oscillator}(c)
($p=0.9\Delta$) and  \ref{fig-single-oscillator}(d) ($p=3\Delta$),
where filled and open circles represent stable and unstable fixed points, respectively.
(Such figures are called phase portraits \cite{Strogatz}.)
In the above simulation,
the trajectory is initially a small closed orbit around the origin.
As $p$ is increased slowly,
the orbit changes while keeping its area $\int dx dy$ constant according to 
the adiabatic theorem in classical mechanics,
where the area is called adiabatic invariance \cite{Landau}.
Thus, above the bifurcation point ($p>\Delta$),
the orbit moves to one of the stable fixed points 
(local minima of the energy surface).

Here we move on to the quantum dynamics of the KPO.
We numerically solved the Schr\"odinger equation
with $H_1$,
where the Hilbert space was truncated at a ``photon" number of 20,
the initial state was set to the ``vacuum" $|0\rangle$,
and $p$ was increased linearly from zero as in the above classical simulation.
Figures  \ref{fig-single-oscillator}(e) and \ref{fig-single-oscillator}(f)
show the Wigner function $W(x,y)$
at $p=0.9\Delta$ and $p=5\Delta$, respectively,
where the Wigner function is a quasiprobability distribution for 
quadrature amplitudes \cite{Leonhardt} (also see Appendix \ref{sec-Wigner}).
In Figs. \ref{fig-single-oscillator}(e) and \ref{fig-single-oscillator}(f),
filled and open circles represent classical stable and unstable fixed points, respectively,
as in Figs. \ref{fig-single-oscillator}(c) and \ref{fig-single-oscillator}(d).
Note that the peaks of the Wigner function are at the classical stable fixed points.
The Wigner distributions show that the stable fixed points are 
accompanied by quantum fluctuations due to the Heisenberg uncertainty relation
for $(a+a^{\dagger})/2$ and 
$(a-a^{\dagger})/2i$. 
On the other hand,
the interference fringe between the two peaks in Fig. \ref{fig-single-oscillator}(f)
means that the two oscillation states are superposed,
that is, a cat state is generated \cite{Leonhardt}.
These results suggest that 
while the classical system chooses one of the two stable branches 
(which branch the system will choose may be unpredictable because of chaos), 
the quantum system can follow both the branches ``simultaneously" 
as a superposition of two coherent states corresponding to the two branches. 
(A coherent state $|\alpha \rangle$ is defined as the eigenstate of $a$:
$a |\alpha \rangle = \alpha |\alpha \rangle$
\cite{Leonhardt}.)
To emphasize this nonclassical feature of a quantum nonlinear oscillator, 
we refer to such an intriguing process as a \textit{quantum-mechanical bifurcation}.

The cat-state generation is explained by the quantum adiabatic theorem
\cite{Messiah} as follows.
The initial state $|0\rangle$ is the ground state for $H_1$ with $p=0$.
As $p$ is increased slowly, the system follows adiabatically 
the ground state of $H_1(t)$.
Finally, $p$ becomes much larger than $\Delta$ and 
the final state becomes approximately the ground state of $H_1$ with $\Delta=0$,
which is a superposition of two coherent states $|\pm \sqrt{p/K} \rangle$.
Since $H_1$ is symmetric under parity inversion $a\to -a$,
the final state should have the same parity as $|0\rangle$.
Consequently, the final state is approximately the cat state
$\displaystyle \left(|\sqrt{p/K} \rangle + |-\sqrt{p/K} \rangle \right)/\sqrt{2}$,
where $\langle \sqrt{p/K} | -\sqrt{p/K} \rangle = e^{-2p/K}$
has been ignored assuming sufficiently large $p$.

Recently, deterministic cat-state generation has been demonstrated
in two different ways with superconducting microwave resonators coupled to a superconducting artificial atom
\cite{Leghtas2015a,Vlastakis2013a}.
The above result provides another method for deterministic cat-state generation
based on quantum adiabatic evolution.
To realize the cat-state generation, 
we require a large Kerr coefficient compared to a loss rate. 
While this requirement is too stringent for optical and mechanical systems, 
superconducting circuits with Josephson junctions have already achieved this situation
\cite{Kirchmair2013a,Rehak2014a}. 
Thus, superconducting systems are most promising for implementing the present scheme.

Next, our quantum computer with KPOs is described. 
If we have $N$ independent KPOs, 
we will obtain a superposition of $2^N$ oscillation states 
via the quantum-mechanical bifurcation described above. 
To exploit the exponentially large number of states 
for solving combinatorial optimization problems, 
we couple the KPOs to one another appropriately depending on given problems.

The combinatorial optimization problem studied here is the Ising problem: 
given a dimensionless Ising energy 
\begin{align}
E_{\tiny{\mbox{Ising}}} = -\frac{1}{2} \sum_{i=1}^N \sum_{j=1}^{N} J_{i,j} s_i s_j,
\label{eq-EIsing}
\end{align}
we want to find a spin configuration minimizing $E_{\tiny{\mbox{Ising}}}$.
Here the Ising spin $s_i$ takes $\pm 1$, $N$ is the number of spins, and
the coupling coefficients satisfy $J_{i,i}=0$ and $J_{i,j}=J_{j,i}$.
Here note that two configurations $\{ s_i \}$ and $\{ -s_i \}$ 
give the same Ising energy, 
and therefore there are always two solutions for each problem. 
The Ising problem is extremely hard unless the coupling topology is too simple;
more precisely,
it is known to be non-deterministic polynomial-time hard (NP-hard)
in computational complexity theory \cite{Barahona1982a}.
Recently, machines specially designed for the Ising problem
have attracted much attention 
\cite{Johnson2011a,Lanting2014a,Ronnow2014a,Boixo2014a,Heim2015a,Utsunomiya2011a,Wang2013a,Marandi2014a}.

For the above problem, we couple $N$ KPOs as follows:
\begin{align}
H=\sum_{i=1}^N H_1^{(i)}
- \hbar \frac{\xi_0}{2} \sum_{i=1}^N \sum_{j=1}^{N} J_{i,j} 
\left( a^{\dagger}_i a_j + a_i a^{\dagger}_j \right),
\label{eq-total-Hamiltonian}
\end{align}
where $H_1^{(i)}$ is the Hamiltonian for the $i$th KPO 
of the form of Eq. (\ref{eq-Hamiltonian1}) 
with an individually controllable detuning $\Delta_i$
and $\xi_0$ is a positive constant with the dimension of frequency.
Note that the coupling Hamiltonian describes standard linear couplings, and therefore
is physically feasible.
It is also notable that $H$ is symmetric under \textit{simultaneous} parity inversion 
defined as $a_i \to -a_i$ for all $i$ simultaneously. 
In the following, we show that the KPO network can solve the Ising problem 
via quantum adiabatic evolution.

To use a quantum adiabatic evolution for finding a configuration
minimizing $E_{\tiny{\mbox{Ising}}}$,
the initial state $|0\rangle$ should be the ground state of $H$ with $p=0$.
This condition can be satisfied by setting the detunings 
such that the following matrix $M$ becomes positive semidefinite
(see Appendix \ref{sec-proof} for the proof):
\begin{align}
M_{i,i}= \Delta_i, \quad
M_{i,j}=-\xi_0 J_{i,j}~(i\neq j).
\label{eq-M}
\end{align}
A simple setting satisfying this condition is as follows
(see Appendix \ref{sec-proof}):
\begin{align}
\Delta_i = \xi_0 \sum_{j=1}^N |J_{i,j}|.
\label{eq-detuning-condition}
\end{align}

By increasing $p$ slowly,
we obtain the ground state of $H$ with large $p$ 
assuming that the so-called adiabatic condition
\cite{Farhi2000a,Das2008a,Messiah} is satisfied.

When $p$ becomes much larger than $|\Delta_i|$ and $\xi_0 |J_{i,j}|$, 
the nonlinear terms in $H$ are dominant, 
the ground state of which is $2^N$-fold degenerate and 
the eigenspace is spanned by the tensor products of $|\pm \sqrt{p/K} \rangle$. 
By the perturbation theory to the lowest order
\cite{Messiah}, 
the correction to the energy of a tensor product 
$|s_1 \sqrt{p/K} \rangle \cdots |s_N \sqrt{p/K} \rangle$ ($s_i= \pm 1$) is given by 
\begin{align}
E_{corr}
=
\hbar \frac{p}{K}
\sum_{i=1}^N \Delta_i
- \hbar \frac{p}{K} \xi_0 \sum_{i=1}^N \sum_{j=1}^{N} J_{i,j} s_i s_j,
\label{eq-energy-correction}
\end{align}
where $\langle \sqrt{p/K} | -\sqrt{p/K} \rangle = e^{-2p/K}$
has been ignored assuming sufficiently large $p$.
Note that the first term in the right-hand side of Eq. (\ref{eq-energy-correction}) 
is independent of $\{ s_i \}$ and the second one is proportional to the Ising energy. 
Consequently, the ground state is two-fold degenerate and the eigenspace is spanned by 
$|S_1 \sqrt{p/K} \rangle \cdots |S_N \sqrt{p/K} \rangle$ and 
$|-S_1 \sqrt{p/K} \rangle \cdots |-S_N \sqrt{p/K} \rangle$, 
where $\{ S_i \}$ and $\{ -S_i \}$ are the two solutions of the Ising problem. 
The degeneracy comes from the simultaneous parity symmetry of $H$. 
Taking the simultaneous parity symmetry of $H$ into account, 
the ground state obtained as a final state in the above adiabatic evolution is given by
\begin{widetext}
\begin{align}
|\psi_f \rangle = 
\frac{|S_1 \sqrt{p/K} \rangle \cdots |S_N \sqrt{p/K} \rangle
+ |-S_1 \sqrt{p/K} \rangle \cdots |-S_N \sqrt{p/K} \rangle}{\sqrt{2 (1+e^{-2Np/K})}}.
\label{eq-psi_f}
\end{align}
\end{widetext}
Thus, it turns out that we can find a solution of the Ising problem 
by measuring the amplitudes of the KPOs and identifying their signs with the Ising spins.

The above degeneracy means that the energy gap 
between the ground state and the first excited state will vanish, 
which seems problematic for the adiabatic approach. 
However, this causes no problem for the following two reasons. 
First, the transition between the two states is prohibited by the simultaneous parity symmetry of $H$. 
(The ground and first excited states have even and odd parities, respectively.) 
Second, even if the transition occurs by some accidental errors, 
we can find a solution correctly because the first excited state is also a superposition of 
$|S_1 \sqrt{p/K} \rangle \cdots |S_N \sqrt{p/K} \rangle$ and 
$|-S_1 \sqrt{p/K} \rangle \cdots |-S_N \sqrt{p/K} \rangle$
(the same as $|\psi_f \rangle$ except for a negative relative phase).

Here it is notable that an entangled cat state given by Eq. (\ref{eq-psi_f}) 
is generated as a result of the quantum computation. 
We confirmed this fact by numerical simulation 
in the case of two spins (see Appendix \ref{sec-two-spin-simulation}). 
Thus, the present scheme also provides a method 
for the generation of the intriguing states via quantum adiabatic evolution.

Finally, we present numerical simulation results of the quantum computation for four-spin problems, 
which are more difficult than two- and three-spin ones in the sense that, in the four-spin case, 
there may be not only frustration but also a nonglobal local minimum. 
In these simulations, the Schr\"odinger equation with $H$ in Eq. (\ref{eq-total-Hamiltonian})
is numerically solved, 
where the Hilbert space is truncated at a ``photon" number of 18 for each KPO, 
the initial state is set to $|0\rangle$, $\xi_0=0.5K$, 
the detunings are set as in Eq. (\ref{eq-detuning-condition}), 
and $p$ is increased linearly from $p(0)=0$ to $p(700/K)=7K$.

\begin{figure}[htbp]
	\includegraphics[width=8.5cm]{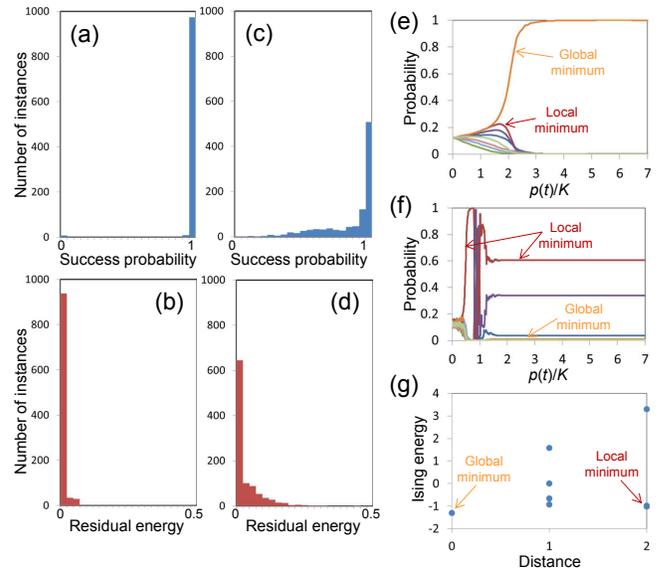}
	\caption{Simulation results for four-spin Ising problems. 
	(a) and (b) show the histograms for the success probabilities and the residual energies, 
	respectively, in the quantum computation estimated by numerically 
	solving the Schr\"odinger equation with $H$ in Eq. (\ref{eq-total-Hamiltonian}). 
	(c) and (d) show the similar results for the classical model. 
	(e) and (f) show the time evolutions of the probabilities of the spin configurations 
	in the quantum and classical models, respectively, 
	for one of the most difficult instances, 
	for which the classical model almost always fails. 
	The energy landscape of this instance is shown in (g), 
	where the distance is defined 
	by Eq. (\ref{eq-distance}). 
	Here two configurations $\{ s_i \}$ and $\{ -s_i \}$ 
	are treated as a pair because these give 
	the same value of $E_{\tiny{\mbox{Ising}}}$.}
	\label{fig-four-oscillator}
\end{figure}

We generated 1000 instances of the problem with the coupling coefficients 
chosen randomly in the range of $-1$ to 1. 
We estimated the success probability and the residual energy 
for each instance [see Appendix \ref{sec-success-probability} for details], 
where the residual energy is defined as 
the difference between the Ising energy obtained 
by simulation and its minimum value \cite{Santoro2002a}. 
The histograms of the success probabilities and the residual energies are shown 
in Figs. \ref{fig-four-oscillator}(a) and \ref{fig-four-oscillator}(b), respectively. 
In Fig. \ref{fig-four-oscillator}, we treat two configurations $\{ s_i \}$ 
and $\{ -s_i \}$ as a pair 
because these give the same value of $E_{\tiny{\mbox{Ising}}}$.

We also simulated a classical model for the quantum computation, 
the equations of which are derived in a similar manner to 
Eqs. (\ref{eq-classical-KPO-x}) and (\ref{eq-classical-KPO-y}) 
(see Appendix \ref{sec-classical}). 
From the results for a single KPO, 
the comparison between the quantum and classical models 
may be helpful for understanding the simulation results. 
For each instance, we repeated the simulation $10^3$ times,
setting the initial values of $x_i$ and $y_i$ to random numbers
in the range of $-10^{-6}$ to $10^{-6}$.
The success probability and the residual energy for each instance
are estimated by taking averages. 
The results are shown in Figs. \ref{fig-four-oscillator}(c) and \ref{fig-four-oscillator}(d).

First, it is notable that 
the classical model can find optimal solutions with high probability. 
This result comes from the fact that the classical model can approximately solve the Ising problem 
(see Appendix \ref{sec-classical}). 
The high success probability for the classical model means 
that the approximation is fairly good. 
(This model may provide a new approach to combinatorial optimization problems 
with a nonlinear system, which may exhibit chaotic behaviors.)

Next, it is clear that 
the quantum model can achieve higher performance than the classical one
for both the success probability and the residual energy.
Since the differences between the two models may be mainly 
quantum superposition and quantum fluctuation, 
the high performance may come from these quantum effects. 
To examine this point, 
we look into one of the most difficult instances, 
for which the classical model almost always fails.

The time evolutions of the probabilities for the spin configurations 
in the quantum and classical models are shown 
in Figs. \ref{fig-four-oscillator}(e) and \ref{fig-four-oscillator}(f), respectively.
Figure \ref{fig-four-oscillator}(g) shows the energy landscape of this instance,
where the distance between a configuration $\{ s_i \}$ and the solution $\{ S_i \}$ is defined as
\begin{align}
D(\{ s_i \},\{ S_i \})
=\min \left( \sum_{i=1}^4 |s_i - S_i|/2 , \sum_{i=1}^4 |s_i + S_i|/2 \right).
\label{eq-distance}
\end{align}

In this problem, there are two local minima and the eigenvector of $M$ for the smallest eigenvalue
corresponds to the nonglobal one.
Thus, the classical model is trapped around the local minimum [Fig. \ref{fig-four-oscillator}(f)]
(see Appendix \ref{sec-classical}).
On the other hand,
in the quantum model,
a superposition of the two local minima arises from quantum fluctuations, 
and finally the probability for the global minimum converges to unity via quantum adiabatic evolution
[Fig. \ref{fig-four-oscillator}(e)]. 
From this result, 
we conclude that quantum superposition and quantum fluctuation work effectively 
to find optimal solutions in the present computation. 
(A more detailed comparison between the two models 
may become a new theme in the field of quantum chaos, 
which is beyond the scope of the present work.)

In conclusion,
we have proposed a quantum computer comprising
quantum nonlinear oscillators exhibiting a quantum-mechanical bifurcation.
The quantum computer solves combinatorial optimization problems
via quantum adiabatic evolution,
where nonlinear terms are increased slowly.
Since this mechanism is different from 
that of conventional adiabatic quantum computation or quantum annealing,
where quantum fluctuation terms are decreased slowly,
we refer to the present approach as
bifurcation-based adiabatic quantum computation.
By simulating four-spin Ising problems and comparing the results with 
those for a classical model,
we have concluded that quantum superposition and quantum fluctuation
work effectively to find optimal solutions.
While conventional quantum computers with qubits and quantum gates \cite{Nielsen,Ladd2010a} 
are analogous to current digital computers with bits and logic gates, 
the present one is analogous to neural computers \cite{Mackay}, 
which are also networks of nonlinear components. 
Thus, the present scheme will lead to the emergence of a new paradigm, 
which may be called ``quantum soft computing," 
 in the fields of quantum information science, nonlinear science, and artificial intelligence.

\section*{Acknowledgments}

I acknowledge Koichi Mizushima for useful comments.

\begin{appendix}

\section{The Wigner function}
\label{sec-Wigner}

The Wigner function $W(x,y)$ is a quasiprobability distribution for quadrature amplitudes, 
$x$ and $y$, defined in the case of a single oscillator as follows \cite{Leonhardt}:
\begin{align}
W(x,y)
=
\frac{1}{\pi}
\int_{-\infty}^{\infty}
e^{2iyz}
\left\langle x - \frac{z}{2} \right| \rho \left| x+ \frac{z}{2} \right\rangle dz,
\label{eq-Wigner}
\end{align}
where $\rho$ is the density operator for the system and 
$|x\rangle$ denotes the eigenstate for $(a+a^{\dagger})/2$. 
The Wigner function has a useful property that 
the integration with respect to $y$ from $-\infty$ to $\infty$ 
gives the probability distribution for $x$, and vice versa. 
Although in the following, we focus on the case of a single oscillator, 
the generalization to multiple oscillators is straightforward.

Using the displacement operator $D(\alpha )=e^{\alpha a^{\dagger} - \alpha^* a}$ and 
the parity operator $P=e^{i \pi a^{\dagger} a}$, 
the Wigner function is rewritten as follows \cite{Leonhardt}:
\begin{align}
W(x,y)
=
\frac{2}{\pi}
\mbox{tr}
\left[
D(2(x+iy))P \rho
\right].
\label{eq-Wigner2}
\end{align}
From this formula, the Wigner function in the number representation is given by
\begin{align}
W(x,y)
=
\frac{2}{\pi}
\sum_{m=0}^{\infty} 
\sum_{n=0}^{\infty} 
(-1)^n
D_{m,n} (2(x+iy)) \rho_{n,m},
\label{eq-Wigner3}
\end{align}
where $\rho_{n,m}$ and $D_{m,n} (\alpha )$ are matrix elements of $\rho$ and $D(\alpha )$, 
respectively, in the number representation. 
$D_{m,n} (\alpha )$ can be expressed as 
\begin{align}
D_{m,n} (\alpha )
=
e^{-|\alpha |^2/2} \sqrt{m! n!}
\sum_{k=0}^{\min (m,n)}
\frac{1}{k!}
\frac{\alpha^{m-k}}{(m-k)!}
\frac{(-\alpha^*)^{n-k}}{(n-k)!}.
\label{eq-Dmn}
\end{align}
We used Eqs. (\ref{eq-Wigner3}) and (\ref{eq-Dmn}) 
for Figs. \ref{fig-single-oscillator}(e) and \ref{fig-single-oscillator}(f).

\section{Calculations of success probabilities and residual energies}
\label{sec-success-probability}

The success probability in Fig. \ref{fig-four-oscillator}(a) is defined 
as the probability that the spin configurations corresponding to the solution are obtained 
in the final measurement of the signs of the quadrature amplitudes. 
Here we explain how to calculate the probability that a spin configuration is obtained. 
The residual energies in Fig. \ref{fig-four-oscillator}(b) are 
obtained by taking expectation values with the probabilities. 
Although in the following, we focus on the case of a single oscillator, 
the generalization to multiple oscillators is straightforward.

\begin{widetext}

The probability that a positive quadrature amplitude is obtained is 
given with the Wigner function as follows:
\begin{align}
P_+
=
\int_0^{\infty} dx \int_{-\infty}^{\infty} dy W(x,y)
=
\int_0^{\infty} rdr \int_{-\frac{\pi}{2}}^{\frac{\pi}{2}} d\phi W(r,\phi),
\label{eq-P+}
\end{align}
where we have used the polar coordinates. 
Using Eqs. (\ref{eq-Wigner3}) and (\ref{eq-Dmn}), we obtain
\begin{align}
P_+
=
\sum_{m=0}^{\infty} 
\sum_{n=0}^{\infty}
I_{m,n}^{(+)}
\mbox{Re} (\rho_{n,m}),
\nonumber
\end{align}
where
\begin{align}
I_{m,n}^{(+)}
=
\left\{
\begin{array}{lll}
\displaystyle
\frac{n!}{2}  
\sum_{k=0}^n
\frac{(-1)^k}{k!} \frac{2^{n-k}}{(n-k)!}
&
\cdots
&
m=n
\\
0 &\cdots & m\neq n \wedge m \equiv n~ (\mbox{mod}~2)
\\
\displaystyle
\sqrt{\frac{m! n!}{2\pi}} 
\frac{(-1)^{\frac{m-n-1}{2}}}{m-n}
\sum_{k=0}^{\min (m,n)} 
\frac{(-1)^k}{k!}
\frac{(m+n-2k)!!}{(m-k)! (n-k)!}
&
\cdots
&
m-n \equiv 1~ (\mbox{mod}~2)
\end{array}
\right.
\nonumber
\end{align}
Similarly, the probability that a negative quadrature amplitude is obtained is given by
\begin{align}
P_-
=
\sum_{m=0}^{\infty} 
\sum_{n=0}^{\infty}
I_{m,n}^{(-)}
\mbox{Re} (\rho_{n,m}),
\nonumber
\end{align}
where
\begin{align}
I_{m,n}^{(-)}
=
\left\{
\begin{array}{lll}
\displaystyle
\frac{n!}{2}  
\sum_{k=0}^n
\frac{(-1)^k}{k!} \frac{2^{n-k}}{(n-k)!}
&
\cdots
&
m=n
\\
0 &\cdots & m\neq n \wedge m \equiv n~ (\mbox{mod}~2)
\\
\displaystyle
\sqrt{\frac{m! n!}{2\pi}} 
\frac{(-1)^{\frac{m-n+1}{2}}}{m-n}
\sum_{k=0}^{\min (m,n)} 
\frac{(-1)^k}{k!}
\frac{(m+n-2k)!!}{(m-k)! (n-k)!}
&
\cdots
&
m-n \equiv 1~ (\mbox{mod}~2)
\end{array}
\right.
\nonumber
\end{align}

\end{widetext}

\section{Proof of the condition for quantum adiabatic evolution}
\label{sec-proof}

Here we prove that
a sufficient condition for that $|0\rangle$ is the ground state for $H$ with $p=0$
is that $M$ defined by Eq. (\ref{eq-M}) becomes positive semidefinite.

The total Hamiltonian
$H$ with $p=0$ can be written as
\begin{align}
H=\hbar \frac{K}{2} \sum_{i=1}^N a_i^{\dagger 2} a_i^2
+ \hbar \sum_{i=1}^N \sum_{j=1}^{N} M_{i,j}  a^{\dagger}_i a_j.
\label{eq-H-appendix}
\end{align}
Since $H|0\rangle =0$,
it is sufficient to show that $H$ is nonnegative.

Since $M$ is a Hermitian matrix,
$M$ is diagonalized as $D=UMU^{\dagger}$,
where $D$ and $U$ are a diagonal matrix and a unitary matrix,
respectively.
Thus we obtain
\begin{align}
H=\hbar \frac{K}{2} \sum_{i=1}^N a_i^{\dagger 2} a_i^2
+ \hbar \sum_{i=1}^N D_{i,i}  b^{\dagger}_i b_i,
\label{eq-H-appendix2}
\end{align}
where $b_i = \sum_{j=1}^N U_{i,j} a_j$.
The operator of this form
is nonnegative when $K$ and all $D_{i,i}$ are nonnegative.
All $D_{i,i}$ are nonnegative by the assumption that $M$ is positive semidefinite. 
Thus, the proof is completed.

This condition is satisfied by choosing $\Delta_i$ as Eq. (\ref{eq-detuning-condition}).
This is easily confirmed as follows.
Consider the following quadratic form of real variables $\{ \eta_i \}$:
\begin{align}
\sum_{i=1}^N \sum_{j=1}^{N} M_{i,j} \eta_i \eta_j
=
\xi_0
\sum_{i=1}^N \sum_{j=i+1}^{N} 
|J_{i,j}|
\left(
\eta_i
-
\frac{J_{i,j}}{|J_{i,j}|}  \eta_j
\right)^2,
\nonumber
\end{align}
where Eq. (\ref{eq-detuning-condition}) has been used.
This quadratic form is always nonnegative.
Therefore, $M$ is positive semidefinite.

\section{Simulation of the quantum computation for a two-spin Ising problem with a ferromagnetic coupling}
\label{sec-two-spin-simulation}

As a simplest problem, 
we considered a two-spin problem with a ferromagnetic coupling: $J_{1,2}=J_{2,1}=1$. 
In this case, the answer is easy: $s_1=s_2=\pm 1$. 
From Eq. (\ref{eq-psi_f}), 
we will obtain an entangled cat state: 
\begin{align}
|ECS (p) \rangle =
\frac{|\sqrt{p/K} \rangle |\sqrt{p/K} \rangle
+|- \sqrt{p/K} \rangle |- \sqrt{p/K} \rangle}{\sqrt{2(1+e^{-4p/K})}}.
\label{eq-ECS}
\end{align}

\begin{figure}[htbp]
	\includegraphics[width=7cm]{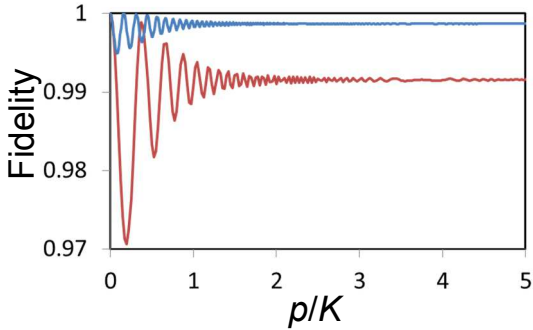}
	\caption{Simulation result of the quantum computation 
	for a two-spin Ising problem with a ferromagnetic coupling: $J_{1,2}=J_{2,1}=1$.
	The fidelity is defined as $F=|\langle ECS (p(t))|\psi (t) \rangle|^2$,
	where $|\psi (t) \rangle$ is the calculated state and 
	$|ECS (p(t)) \rangle$ is the entangled cat state defined by Eq. (\ref{eq-ECS}).
	The parameters are set as $\Delta_1 = \Delta_2 = \xi_0=0.5K$.
	$p$ is increased linearly from zero to $5K$.
	The upper and lower curves correspond to
	the computation times of $500/K$ and $200/K$, respectively.}
	\label{fig-two-oscillator}
\end{figure}

We numerically solved the Schr\"odinger equation with $H$ in Eq. (\ref{eq-total-Hamiltonian}), 
where the Hilbert space was truncated at a ``photon" number of 20 for each KPO, 
the initial state was set to $|0\rangle$, $\Delta_1=\Delta_2=\xi_0=0.5K$, 
and $p$ is increased linearly from zero to $5K$. 
Figure \ref{fig-two-oscillator} shows the fidelity 
between the calculated state $|\psi (t) \rangle$ and 
the entangled cat state $|ECS(p(t)) \rangle$ defined as 
$F=|\langle ECS (p(t))|\psi (t) \rangle|^2$. 
In Fig. \ref{fig-two-oscillator}, 
the upper and lower curves are the results 
for the computation times of $500/K$ and $200/K$, respectively.

The high fidelities in Figs. \ref{fig-two-oscillator} prove 
that the entangled cat state can be generated indeed. 
Thus, the present quantum computation provides 
a simple method for deterministic generation of 
such an intriguing quantum state via quantum adiabatic evolution. 
Figure \ref{fig-two-oscillator} also shows that the more slowly, 
$p$ is increased, the higher the fidelity becomes. 
This is the feature of quantum adiabatic evolution.

\section{Approximate solution by the classical model}
\label{sec-classical}

The equations of motion in the classical model for the present quantum computation are given by
\begin{align}
\dot{x}_i &= y_i \left[ \Delta_i + p + K \left( x_i^2+y_i^2 \right) \right] 
-\xi_0 \sum_{j=1}^N J_{i,j} y_j,
\label{eq-classical-four-KPO-x}
\\
\dot{y}_i &= x_i \left[ -\Delta_i + p - K \left( x_i^2+y_i^2 \right) \right]
+\xi_0 \sum_{j=1}^N J_{i,j} x_j.
\label{eq-classical-four-KPO-y}
\end{align}

Here we show that the classical model can obtain solutions for a relaxation problem of the Ising problem, 
where the relaxation is to replace the Ising spin $s_i$ with 
a continuous variable $\zeta_i$.
We obtain an approximate solution for 
the original problem by identifying the sign of $\zeta_i$ as $s_i$.

First, we restate the Ising problem as 
the following energy is to be minimized:
\begin{align}
E'_{Ising} 
&= -\xi_0 \sum_{i=1}^N \sum_{j=1}^{N} J_{i,j} s_i s_j
+ \sum_{i=1}^N \Delta_i s_i^2
\nonumber
\\
&= \sum_{i=1}^N \sum_{j=1}^{N} M_{i,j} s_i s_j.
\nonumber
\end{align}
Since $s_i^2=1$, this problem is equivalent to the original one.

The relaxation problem is defined as 
the following energy is to be minimized
under the condition that $\sum_{i=1}^N \zeta^2_i =N$:
\begin{align}
E_c
=
\sum_{i=1}^N \sum_{j=1}^{N} M_{i,j} \zeta_i \zeta_j,
\nonumber
\end{align}
where the constraint condition is necessary 
to find nontrivial solutions.
Since $M$ is positive semidefinite,
the solution of the relaxation problem is given by
the eigenvector of $M$ for the smallest eigenvalue.
 Here it is also important that 
 we can obtain a lower bound for the Ising energy 
 from the solution of the relaxation problem.

On the other hand,
the classical model can find such a vector at the first bifurcation point.
Near the bifurcation point,
the quadrature amplitudes are small.
Neglecting the nonlinear terms,
we obtain the following condition for the fixed points:
\begin{align}
y_i \left( \Delta_i + p \right) -\xi_0 \sum_{j=1}^N J_{i,j} y_j 
&= p y_i + \sum_{j=1}^N M_{i,j} y_j = 0,
\nonumber
\\
x_i \left( -\Delta_i + p \right)+\xi_0 \sum_{j=1}^N J_{i,j} x_j 
&= p x_i - \sum_{j=1}^N M_{i,j} x_j = 0.
\nonumber
\end{align}
These are characteristic equations for $M$.
Since $M$ is positive semidefinite and $p \ge 0$,
$\{ y_i \}$ cannot have nontrivial solutions.
On the other hand,
$\{ x_i \}$ has a nontrivial solution at the bifurcation point,
where $p$ is the smallest eigenvalue of $M$.
Then $\{ x_i \}$ is the corresponding eigenvector.

Thus the classical model can find 
the solution for the relaxation problem of the Ising problem.
This may be the reason 
why the classical model can find optimal solutions with high probability, 
as shown in Fig. \ref{fig-four-oscillator}(c).

\end{appendix}


\begin{thebibliography}{19}


\bibitem{Strogatz}
S. H. Strogatz, 
\textit{Nonlinear dynamics and chaos} 
(Westview Press, Boulder, CO, ed. 2, 2015). 




\bibitem{Dykman}
M. Dykman, Ed., 
\textit{Fluctuating nonlinear oscillators} 
(Oxford Univ. Press, Oxford, 2012). 


\bibitem{Leghtas2015a}
Z. Leghtas \textit{et al}.,
Confining the state of light to a quantum manifold by engineered two-photon loss,
Science \textbf{347}, 853-857 (2015).


\bibitem{Mirrahimi2014a}
M. Mirrahimi, \textit{et al}., 
Dynamically protected cat-qubits: a new paradigm for universal quantum computation, 
New. J. Phys. \textbf{16}, 045014 (2014).




\bibitem{Farhi2000a}
E. Farhi, J. Goldstone, S. Gutmann, and M. Sipser, 
Quantum computation by adiabatic evolution,
arXiv:quant-ph/0001106 (2000).


\bibitem{Farhi2001a}
E. Farhi \textit{et al}.,
A Quantum Adiabatic Evolution Algorithm Applied to Random Instances of an NP-Complete Problem,
Science \textbf{292}, 472-475 (2001).



\bibitem{Kadowaki1998a}
T. Kadowaki, H. Nishimori,
Quantum annealing in the transverse Ising model,
Phys. Rev. E \textbf{58}, 5355-5363 (1998).

\bibitem{Santoro2002a}
G. E. Santoro, R. Marto\v{n}\'ak, E. Tosatti, R. Car,
Theory of Quantum Annealing of an Ising Spin Glass,
Science \textbf{295}, 2427-2430 (2002).


\bibitem{Das2008a}
A. Das, B. K. Chakrabarti,
Colloquium: Quantum annealing and analog quantum computation,
Rev. Mod. Phys. \textbf{80}, 1061-1081 (2008).

\bibitem{Vlastakis2013a}
B. Vlastakis \textit{et al}.,
Deterministically Encoding Quantum Information Using 100-Photon Schr\"odinger Cat States,
Science \textbf{342}, 607-610 (2013).


\bibitem{Kirchmair2013a}
G. Kirchmair, \textit{et al}., 
Observation of quantum state collapse and revival due to the single-photon Kerr effect,
Nature \textbf{495}, 205-209 (2013).

\bibitem{Rehak2014a}
M. Reh\'{a}k, \textit{et al}., 
Parametric amplification by coupled flux qubits,
Appl. Phys. Lett. \textbf{104}, 162604 (2014).



\bibitem{Lin2014a}
Z. R. Lin \textit{et al}.,
Josephson parametric phase-locked oscillator and 
its application to dispersive readout of superconducting qubits,
Nat. Commun. \textbf{5}, 4480 (2014).


\bibitem{Okamoto2013a}
H. Okamoto \textit{et al}.,
Coherent phonon manipulation in coupled mechanical resonators,
Nat. Phys. \textbf{9}, 480-484 (2013).

\bibitem{Faust2013a}
T. Faust, J. Rieger, M. J. Seitner, J. P. Kotthaus, E. M.Weig,
Coherent control of a classical nanomechanical two-level system,
Nat. Phys. \textbf{9}, 485-488 (2013).

\bibitem{Moser2014a}
J. Moser, A. Eichler, J. G\"uttinger, M. I. Dykman, A. Bachtold,
Nanotube mechanical resonators with quality factors of up to 5 million,
Nat. Nanotech. \textbf{9}, 1007-1011 (2014).


\bibitem{Landau}
L. D. Landau and E. M. Lifshits, 
\textit{Mechanics} (Pergamon, Oxford, 3rd ed, 1976).



\bibitem{Leonhardt}
U. Leonhardt,
\textit{Measuring the Quantum State of Light} 
(Cambridge Univ. Press, Cambridge, 1997).


\bibitem{Messiah}
A. Messiah, \textit{Quantum Mechanics} (Wiley, New York, 1976).







\bibitem{Barahona1982a}
F. Barahona,
On the computational complexity of Ising spin glass models,
J. Phys. A \textbf{15}, 3241-3253 (1982).









\bibitem{Johnson2011a}
M. W. Johnson \textit{et al}.,
Quantum annealing with manufactured spins,
Nature \textbf{473}, 194-198 (2011).



\bibitem{Lanting2014a}

T. Lanting \textit{et al}.,
Entanglement in a Quantum Annealing Processor,
Phys. Rev. X \textbf{4}, 021041 (2014).


\bibitem{Ronnow2014a}
T. F. R\o nnow \textit{et al}.,
Defining and detecting quantum speedup,
Science \textbf{345}, 420-424 (2014).

\bibitem{Boixo2014a}
S. Boixo \textit{et al}.,
Evidence for quantum annealing with more than one hundred qubits,
Nat. Phys. \textbf{10}, 218-224 (2014).


\bibitem{Heim2015a}
B. Heim, T. F. R\/onnow, S. V. Isakov, M. Troyer,
Quantum versus classical annealing of Ising spin glasses,
Science \textbf{348}, 215-217 (2015).







\bibitem{Utsunomiya2011a}
S. Utsunomiya, K. Takata, Y. Yamamoto,
Mapping of Ising models onto injection-locked laser systems,
Opt. Exp. \textbf{19}, 18091 (2011).







\bibitem{Wang2013a}
Z. Wang, A. Marandi, K. Wen, R. L. Byer, Y. Yamamoto,
Coherent Ising machine based on degenerate optical parametric oscillators,
Phys. Rev. A \textbf{88}, 063853 (2013).


\bibitem{Marandi2014a}
A. Marandi, Z. Wang, K. Takata, R. L. Byer, Y. Yamamoto,
Network of time-multiplexed optical parametric oscillators as a coherent Ising machine,
Nat. Photon. \textbf{8}, 937-942 (2014).




\bibitem{Nielsen}
M. A. Nielsen, I. L. Chuang, 
\textit{Quantum Computation and Quantum Information} 
(Cambridge Univ. Press, Cambridge, 2000). 

\bibitem{Ladd2010a}
T. D. Ladd \textit{et al}.,
Quantum computers,
Nature \textbf{464}, 45-53 (2010).


\bibitem{Mackay}
D. J. C. MacKay,
\textit{Information Theory, Inference and Learning Algorithms}
(Cambridge Univ. Press, Cambridge, 2003).






\end{thebibliography}
\end{document}